\documentclass[journal,comsoc]{IEEEtran}
\usepackage[caption=false]{subfig}
\usepackage[noadjust]{cite}
\usepackage{float}
\usepackage{graphicx}
\usepackage[normalem]{ulem}
\newcommand\semiHuge{\fontsize{23.72}{27.38}\selectfont}
\usepackage{color}
\usepackage{balance}

\begin{document}

\title{{\semiHuge From Federated to Fog Learning:  Distributed Machine Learning over  Heterogeneous Wireless Networks}}

\author{Seyyedali Hosseinalipour, Christopher G. Brinton,  Vaneet Aggarwal, Huaiyu Dai, and Mung Chiang
\thanks{S. Hosseinalipour, C. G. Brinton, V. Aggarwal, and M. Chiang are with Purdue University, IN, USA e-mail: \{hosseina,cgb,vaneet,chiang\}@purdue.edu. H. Dai is with NC State University, NC, USA e-mail: hdai@ncsu.edu.
}}
\maketitle
\begin{abstract}
Machine learning (ML) tasks are becoming ubiquitous in today's network applications. 
Federated learning has emerged recently as a technique for training ML models at the network edge by leveraging processing capabilities across the nodes that collect the data. There are several challenges with employing conventional federated learning in contemporary networks, due to the significant heterogeneity in compute and communication capabilities that exist across devices. To address this, we advocate a new learning paradigm called \textit{fog learning} which will intelligently distribute ML model training across the continuum of nodes from edge devices to cloud servers. Fog learning enhances federated learning along three major dimensions: \textit{network}, \textit{heterogeneity}, and \textit{proximity}. It considers a multi-layer hybrid learning framework consisting of heterogeneous devices with various proximities. It accounts for the topology structures of the local networks among the heterogeneous nodes at each network layer, orchestrating them for collaborative/cooperative learning through device-to-device (D2D) communications. This migrates from star network topologies used for parameter transfers in federated learning to more distributed topologies at scale. We discuss several open research directions to realizing fog learning.
\end{abstract}


\section{Introduction}\label{sec:intro}
\noindent The modern era has witnessed an explosion in the number of intelligent wireless devices capable of connecting to the Internet and forming ad-hoc networks. The improved processing capabilities of these Internet of Things (IoT) devices coupled with rising user demands for data-intensive, latency-sensitive tasks has motivated \textit{fog computing}. Fog computing is an emerging architecture which aims to orchestrate and manage processing resources across nodes in the cloud-to-things continuum, encompassing the cloud, core, metro, edge, clients, and things~\cite{7901470}. Security and privacy of user data is also an important part of this emerging paradigm~\cite{TobeAdded1}.

Machine learning (ML) has attracted significant recent attention in networking applications, given its potential to provide fast and autonomous decision-making for 5G, 6G, and future wireless technologies~\cite{TobeAdded5}.
ML techniques generally require large datasets for model training, especially in the newer category of deep learning. This data is generated at end user devices as they interact with applications, and then traditionally is transferred to a central datacenter which carries out the model training. Consider, for example, automated facial recognition carried out by social media platforms today: when a user uploads a photo, a prediction is made of who is in the image by applying a model trained over billions of samples at a datacenter. The user's feedback on this prediction (e.g., whether it is correct) informs further model refinement.

Centralized ML model training is prohibitive in many emerging network applications, however. 

In particular, transferring large volumes of data samples from the end users to the cloud has the following drawbacks:
\begin{enumerate}
\item For battery-limited devices such as smartphones, unmanned aerial vehicles (UAVs), and wireless sensors, uplink data offloading can consume prohibitive amounts of energy.
\item For latency-sensitive applications, the round trip time of data transfer, model training/up-dating, and decision making can be prohibitively long.
\item In privacy-sensitive applications, end users may not be willing to share their raw data.
\end{enumerate}
These limitations have motivated work on distributed ML model training, where federated learning has received significant recent attention~\cite{8865093}.

\subsection{Federated Learning}
\begin{figure}[t]
\includegraphics[width=.45\textwidth]{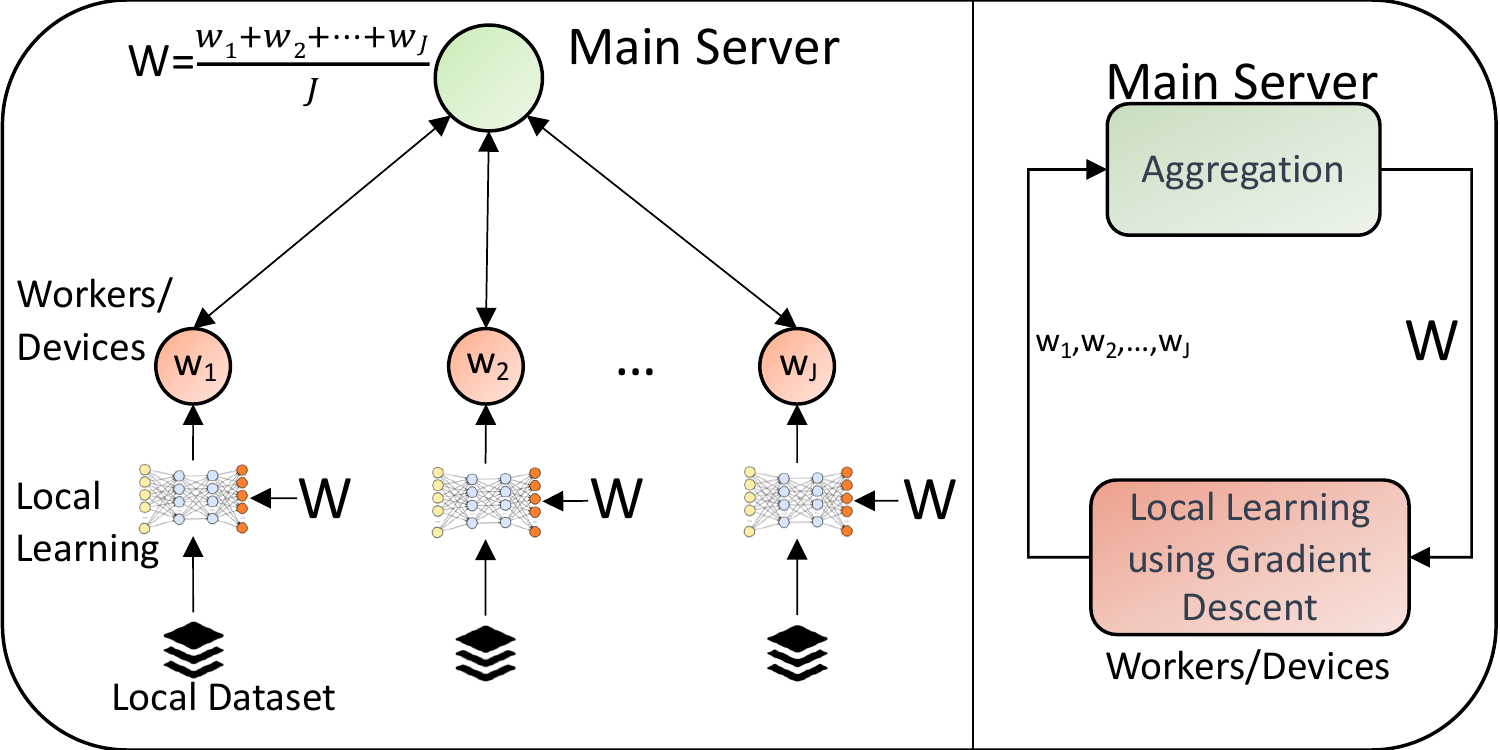}
\centering
\caption{Left: Conventional ``star topology'' of federated learning. Right: An abstract model of data flow in federated learning.}
\label{fig:simpleFL}
\vspace{-0.15in}
\end{figure}

The standard operation of federated learning is depicted in Fig.~\ref{fig:simpleFL}. To train an ML model (e.g., a neural network), two steps are repeated in sequence: (i) \textit{local learning}, in which each worker device updates the parameters of the ML model (e.g., weights on neurons) using its collected dataset, and (ii) \textit{global aggregation}, in which a main server determines the new global model from the local updates and synchronizes the devices with this aggregated version. The local learning at each device typically consists of gradient descent iterations to update the model. The global aggregation is typically an averaging of the local parameters, which may be weighted depending on the perceived quality of devices' updates~\cite{8865093}.
 A key property of federated learning is that the data itself is never transferred between the devices and the server, which further reduces communication demands, and mitigates privacy concerns associated with data sharing.

The standard implementation of federated learning causes performance issues in contemporary fog networking environments, however. Next, we outline the key considerations for developing \textit{network-aware} techniques for distributing ML tasks, and initial works that have attempted to address them.

\subsection{Design Considerations for Network-Aware ML}
\label{ssec:consider}
\subsubsection{Communication heterogeneity}
Most of the IoT devices engaged in ML -- cellular phones, smart vehicles, wireless sensors, UAVs, etc. -- are mobile, with significant heterogeneity in their communication abilities. Channel qualities will change over time and as devices move through the network.
As the achievable uplink and downlink data rates of the system will vary for each node over time, they must be taken into consideration in the design of distributed ML techniques. These heterogeneous communication characteristics have motivated several recent studies on federated learning for wireless networks, e.g.,~\cite{8737464}. Additionally, they have motivated studies on communication-efficient federated learning, through the techniques of quantization (i.e., compressing model updates prior to transmission) and sparsification (i.e., transmitting only some elements of the parameter vectors)~\cite{8865093}.

\subsubsection{Computation/storage heterogeneity}
Wireless devices exhibit heterogeneity in their processing equipment and availability of their resources. Thus, the time required to perform a single local update will vary from one device to another.
This has motivated studying the effects of device compute delays and the existence of stragglers on the time required to train ML models~\cite{8737464}. Methods that have been proposed to resolve these effects mostly rely on intelligent selection of device training participation. Techniques for mitigating compute limitations have also been studied more generally, e.g., through model compression~\cite{wang2018wide}.

\subsubsection{Privacy and security}
Although federated learning eliminates the need to transmit raw data, it is possible for sensitive information to be leaked through reverse engineering of model parameters~\cite{tu2020network}. 
This  has motivated investigations into adapting well-known privacy and security-preservation techniques -- such as differential privacy and functional encryption -- to federated learning~\cite{hardy2017private}.

\subsubsection{Joint performance metrics}
The performance of an ML task is typically measured through the convergence speed and the accuracy of the resulting model. In network-aware ML, the previous three design considerations suggest additional performance metrics. But these objectives tend to compete with one another: for example, a wireless network device processing more gradient updates may improve resulting model quality, but requires more energy consumption.
Thus, techniques for network-aware ML must consider a joint optimization among the objectives of (i) minimizing network resource costs, (ii) maximizing resulting model quality, and (iii) maximizing privacy/security, with different importance assigned to each objective depending on the application~\cite{8664630,tu2020network}.

\subsection{Dimensions of Innovation for Network-Aware ML}
Compared with federated learning, fog learning is defined by the three dimensions of \textit{network}, \textit{heterogeneity}, and \textit{proximity}:
\begin{enumerate}
\item It considers the networks and topology structures among the devices and incorporates a collaboration/cooperation among local wired/wireless nodes using device-to-device communications.
\item It considers the heterogeneity of nodes through the cloud-to-things continuum, in terms of computation capability and local data distributions.
\item It exploits the proximity of resource-limited nodes to resource-abundant nodes to optimize ML training.
\end{enumerate}

\begin{figure*}[t]
\includegraphics[width=0.98\textwidth]{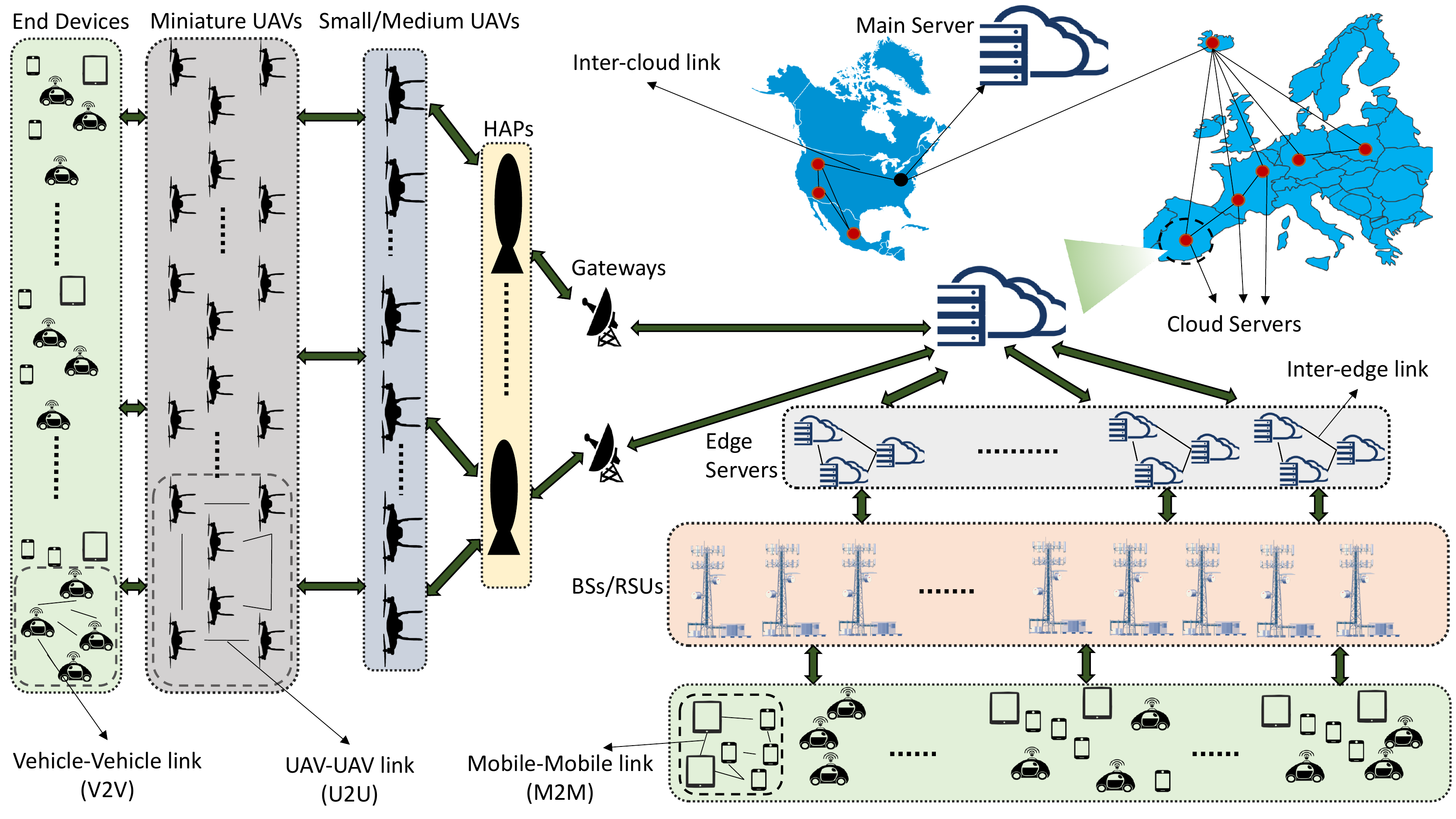}
\centering
\caption{A schematic of model aggregation stages for a large-scale ML task in network-aware learning. The main server aggregates parameter updates from multiple cloud servers. Before reaching these cloud servers, local models trained by devices goes through multiple layers of aggregations. The devices can learn cooperatively via direct D2D communications, through which model parameters, datasets or both are exchanged.}
\label{fig:multilayer}
\vspace{-0.15in}
\end{figure*}

\section{Motivating a New Architecture \\ for Network-Aware Learning}
\noindent Conventional federated learning suffers from a series of limitations when implemented over fog networks. In this section, we will explain these limitations, motivating a new paradigm for distributed ML.

\subsection{Federated Learning: Limitations in Fog Environments}
Consider training and managing a data-intensive, latency-sensitive ML task over a large-scale fog network. We face the following key limitations using federated learning as the solution:

\subsubsection{Multi-layer nature of large-scale learning}
Under federated learning, global aggregations would be performed at the main datacenter. When smartphones, smart vehicles, or other connected edge devices perform their local updates, their cellular base stations (BSs), road side units (RSUs), or analogous access points cannot directly transfer these learned parameters to the main server, which will be in a datacenter located possibly thousands of miles away. Instead, one pragmatic approach would be to consider multiple aggregations at different scales, e.g., edge servers in localities, cities, states, etc., before finally reaching the datacenter. Similarly, for a team of data-gathering UAVs in an area with no cellular coverage, the local learning parameters may first be aggregated by a team of miniature UAVs, then multiple heavier UAVs, and then a high altitude platform (HAP). The HAP would transmit the aggregated models to an edge server through a backhaul network. Once at the edge server, these parameters could traverse the aforementioned hierarchy to reach the main server.
This potential multi-layer network structure for model aggregation is depicted in Fig.~\ref{fig:multilayer}.

\subsubsection{Overloading heterogeneous network resources}
Current cellular BSs and RSUs are not designed to handle model uploads from large numbers of active devices simultaneously. Training deep neural networks (DNNs) with federated learning can require participation from many active devices, as high complexity models require large datasets~\cite{wang2018wide}. Moreover, given the heterogeneity of IoT devices, each participating device may only be capable of processing a small set of samples for a high dimensional model. This calls for a learning architecture that optimizes the choice of devices participating in model uploading based on current network conditions.

\subsubsection{Device collaboration/cooperation} Federated learning ignores the topology structures among the devices and the possibility of collaboration/cooperation among the devices without engaging the main server. Enabling direct communication between devices in local neighborhoods of each network layer could lead to significant power and bandwidth savings by reducing uplink transmissions to nodes in the higher layers. This calls for a framework that explicitly considers device-to-device (D2D) communications being enabled in 5G-and-beyond wireless. We will refer to all communication between devices/nodes within a single network layer as D2D, examples of which are depicted in Fig.~\ref{fig:multilayer}.

\subsubsection{Strict privacy assumptions}
Federated learning guarantees that each device's local dataset is never transferred over the network. While this is important in privacy-sensitive applications, in many cases users may be willing to share portions of their datasets for ML training, which can be useful when there is a combination of resource-hungry and resource-rich devices. For example, a smart car attempting to train an object classifier with a limited on-board processor is likely willing to offload its sensor data to a more computation powerful car to expedite the training process if the channel conditions are reasonable. This calls for a learning framework which can adapt based on privacy needs.

\subsection{From Federated to Fog Learning}
Given these limitations, we propose a new learning paradigm called \textit{fog learning}. As opposed to federated learning which is based on a star topology of device-server interactions, fog learning will explicitly consider the network and topology structures among the devices and enable intelligent device collaborations/cooperations  through data and parameter offloading. This hybrid learning paradigm will exploit the multi-layer structure of fog networks to optimize performance in the presence of heterogeneous network resources. 

There are some recent works on hierarchical federated learning, e.g.,~\cite{TobeAdded3,TobeAdded2}. These works are mainly focused on specific two-tiered network structures above wireless cellular devices, e.g., edge clouds connected to a main server~\cite{TobeAdded3} or small cell and macro cell base stations~\cite{TobeAdded2}. Fog learning generalizes this concept to a multi-layer structure that encompasses all IoT elements between the end devices and the main server. Moreover, fog learning introduces collaborative/cooperative model training via D2D communications among the devices at different layers of the network hierarchy.

\section{Fog Learning: a  Multi-layer \\ Hybrid Learning Paradigm}
\noindent In this section, we define fog learning in terms of its multi-layer structure and hybrid learning characteristics.

\subsection{Multi-layer Network Architecture}
Fog learning is a multi-layer learning architecture over a fog network. Similar to conventional federated learning, the main server will conduct global aggregations. However, the end users are not directly connected to the main server: instead, the local models learned by end devices may traverse multiple layers of aggregations before reaching the main server. Local aggregations at each layer provide dimensionality reduction, reducing the size of the data being transmitted upstream. Synchronizations at each layer also provide agile responses to any changes in local data distributions.

\begin{figure}[t]
\includegraphics[width=.47\textwidth]{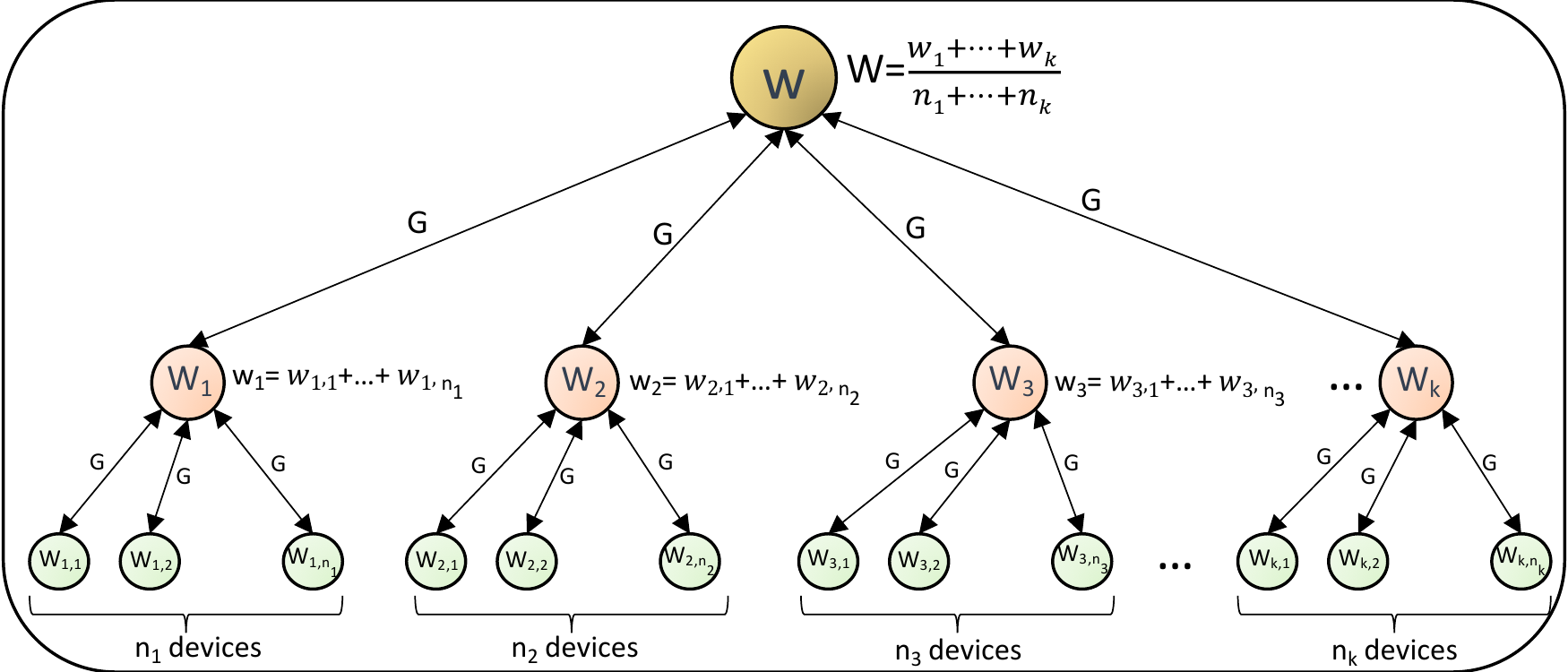}
\centering
\caption{Dimensionality reduction from multi-layer aggregations. The length of the original learning parameter vectors at each end device is $G$. The size of data transmitted upstream from each middle node is also $G$, reduced by a factor of the number of node inputs.} 
\label{fig:aggregate}
\vspace{-0.15in}
\end{figure}

To see the motivation for dimensionality reduction, consider that any ML model is represented as a vector of its model parameters. For a DNN, this vector can have millions of entries~\cite{wang2018wide}, where each element requires a certain number of bits for storage and transfer. Depending on the quantization method, then, this parameter vector could require anywhere from a few megabytes to gigabytes. For the hierarchical network structure depicted in Fig.~\ref{fig:multilayer}, consecutive transmissions of these vectors from millions of edge devices to the main server would lead to large delays, overloaded network infrastructure, and high communication costs.

Each group of devices in Fig.~\ref{fig:multilayer} forms a learning cluster which conducts local aggregations of its internal parameters. After each local aggregation, the size of the resulting vector to be transmitted upstream is the same as any one of the input vectors, as illustrated in Fig.~\ref{fig:aggregate}. For instance, each UAV in Fig.~\ref{fig:multilayer} can aggregate its associated devices' parameters and send the resulting vector to the upper layer. 

\subsection{Hybrid Learning: Vertical and Horizontal Communications}
The learning architecture in Fig. \ref{fig:aggregate} follows a vertical communication structure, where model parameters are passed only upstream and downstream between the network layers. Fog learning takes this one step further to allow for horizontal communications between devices in the same layer.

Peer-to-peer (P2P) networking has been an area of research, offering on-demand establishment of connectivity and eliminating the requirement of a central module to facilitate communication between peers. 5G-and-beyond wireless technologies are enabling D2D communications between wireless nodes, which is motivating P2P intelligence in fog computing~\cite{tu2020network}. There is a well-developed body of literature on D2D communication protocols for MANETs, VANETs, FANETs, and wireless sensor networks.

\begin{figure*}[t]
\includegraphics[width=0.99\textwidth]{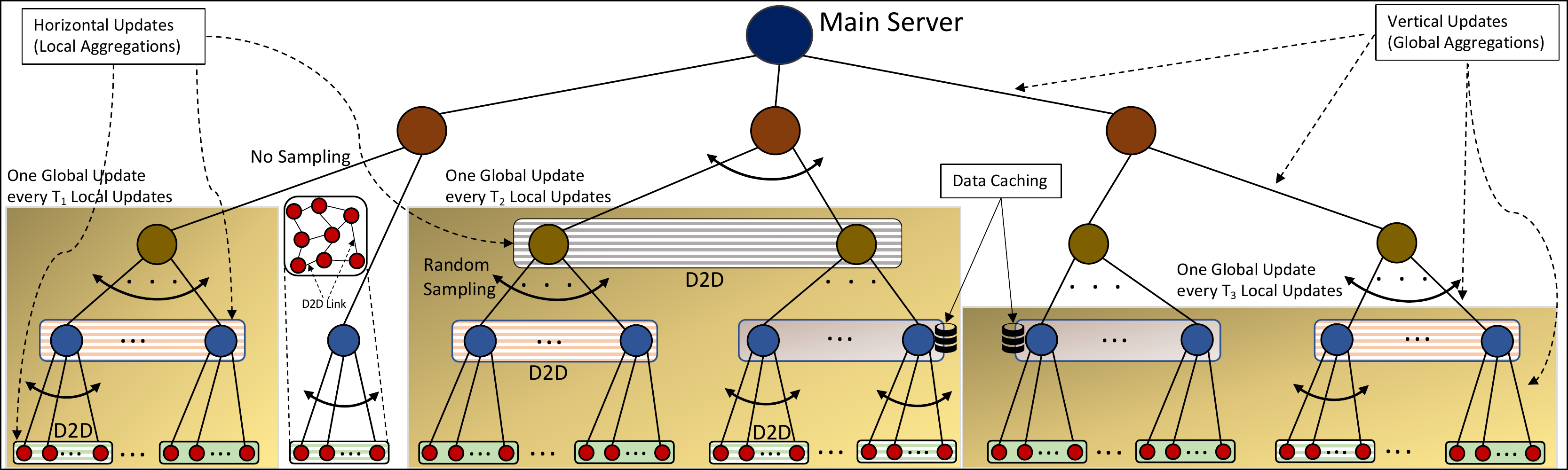}
\centering

\caption{Network representation of fog learning. The root of the tree is the main server, the leaves are the end devices, and the nodes in-between are different intermediate devices. The nodes belonging to the same layer and the same horizontal rectangle form clusters. The patterned rectangles correspond to those clusters that choose to engage in D2D and distributedly learn their model aggregation. The parent nodes of such clusters can then sample one (or a tiny fraction) of their children nodes to obtain the aggregated model. Each yellow block represents a \textit{learning block}, where the top nodes have a certain clock for transmitting model parameters upstream for global aggregations.} 
\label{fig:tree}
\vspace{-0.15in}
\end{figure*}

Considering again the structure in Fig.~\ref{fig:multilayer}, fog learning would intelligently cluster the devices in the bottom-most layer such that each cluster has the potential to form a wireless ad-hoc network for parameter sharing or data offloading. Similarly, the upper layers will be clustered such that the computing nodes in each layer are capable of communicating for parameter sharing, in some cases via low-latency wired connections (e.g., multiple local edge servers connected via fiber in a metropolitan area) and in other cases over the air (e.g., UAVs).

In Fig.~\ref{fig:tree}, we represent the fog learning network architecture as a logical tree graph, the leaves of which are the edge devices and the root of which is the main server. Fog learning is a hybrid learning methodology which leverages horizontal communications among nodes in addition to vertical parameter transfers between the layers. In the following, we first discuss a general approach for D2D communications at different network layers, and then discuss two data offloading strategies that can be utilized in the bottom layers of the network.

\subsubsection{Distributed aggregations through horizontal communications}
\label{sub:genFra}
The nodes inside a D2D-enabled cluster are capable of computing the local aggregation of their locally-trained parameters in a distributed manner, through message passing and consensus formation. This approach eliminates the need for the parent node to compute the aggregation, and can be implemented at all the network layers, which has energy efficiency advantages (discussed further in Sec. \ref{sec:adv}). At the bottom-most layer, the datasets of the devices remain local, as in federated learning. In leveraging such horizontal communications, the conventional star topology used in federated learning is transformed to a collaborative/cooperative distributed fog learning topology.

Our recent work~\cite{TobeAdded8} studied a realization of horizontal aggregation based on a distributed average consensus formation scheme. 
We showed that even with limited amounts of D2D communication enabled, the learning accuracy approaches centralized gradient descent. We demonstrated that using this technique can result in around 50\% device energy savings and 80\% reduction in the number of parameters transferred over the network compared with conventional federated learning.

\subsubsection{D2D offloading under milder privacy concerns}
In addition to sharing learning parameters, the proposed D2D communication scheme can also be used for partial dataset offloading among trusted edge devices, for applications with milder privacy concerns. In Fig.~\ref{fig:tree}, this is only applicable at the bottom-most layer of the tree where the data is collected. This approach is useful in the presence of heterogeneous computation resources within a cluster (discussed further in Sec. \ref{sec:adv}).

Our recent work~\cite{tu2020network} studied the improvement in network resource costs that intelligent D2D data offloading can provide to distributed learning, finding in particular that up to 50\% decrease in the total device processing and transmit resource utilization are possible compared with conventional federated learning. Our results reveal that these gains are consistent over a range of D2D topologies defined by communication restrictions (such as privacy) between nodes.

and found up to 50\% decrease in the total device processing and transmit resource utilization are possible compared with conventional federated learning.

\subsubsection{Inter-layer data offloading and caching}
Mobile devices at the bottom-most network layer may move between local topologies rapidly, which presents an opportunity to improve local data distributions. Specifically, if devices offload portions of non-privacy-sensitive data to the next layer up, this data can be cached and broadcasted among a larger number of edge devices. This will increase the similarity of local data to the global distribution and reduce model bias from local updates.

\subsection{Performance Advantages of Fog Learning}\label{sec:adv}

The local aggregation and D2D offloading features of fog learning will be particularly important for contemporary data-intensive, latency-sensitive applications. These include training ML models for autonomous vehicle navigation, smart factory automation, and augmented/virtual reality (AR/VR) navigation \cite{tu2020network}. Specifically, the advantages provided are as follows:

\subsubsection{Reducing network traffic}
Fog learning employs local aggregations of ML model parameters at different layers of the topology, providing an upstream dimensionality reduction. This results in a significantly reduced network traffic between different network layers. 
Reducing data transfer requirements over long distances decreases latency and communication costs. This is particularly important when training high complexity models like DNNs; in these cases, fog learning can leverage asynchronous layer-wise training and parameter update techniques~\cite{TobeAdded7} for further reductions in upstream traffic.

\subsubsection{Network power savings}\label{sub1}
Horizontal D2D communications allow node clusters to distributedly discover their aggregated models. Thus, the parent node of the cluster can choose one device to upload the aggregated value. Decreasing the number of uplink transmissions by an order of magnitude will reduce energy consumption significantly. For instance, in a cellular network, continuous communication with the BS drains a smartphone's battery rapidly. With D2D enabled, rather than uploading to the BS at each aggregation, the devices could engage in short-range, low power communications, and only one device will need to transmit the result. Instead of selecting one device, it would also be possible to employ a diversity technique where each device in a cluster engages in short, simultaneous uplink transmissions of only a fraction of the parameters.

\subsubsection{Efficient spectrum usage}
Devices in a cluster engaged in D2D communications can operate in out-band mode, which does not require utilizing the licensed spectrum of e.g., a cellular BS or a vehicular RSU. Furthermore, when using in-band D2D, the devices can use opportunistic spectrum access methods to exploit the unused licensed spectrum.

\subsubsection{Adaptation to device mobility}
Devices may enter/exit a local cluster rapidly. When a device enters a D2D enabled cluster, it can join the learning process quickly through acquisition of the current model parameters from a neighboring node. Also, when a device exits, it can transfer its model/data to a local peer so its locally updated model and data is not negated. This capability, along with the fact that devices in different clusters can perform learning in parallel, can be described as \textit{parallel successive learning}: nodes can inherit partially-trained models and continue refining the parameters with newly collected data. 

\subsubsection{Leveraging passive and straggler device datasets}
Certain devices may possess valuable data but may have lower computational capabilities or not be engaged in the training process. With D2D-enabled offloading and active inter-layer data caching, these passive datasets can be transferred to resource-abundant active devices.

\subsubsection{Faster convergence in fewer global aggregations}
\label{sub4}
By mitigating the effect of stragglers and enabling more distributed processing on heterogeneous datasets, the global model in Fig. \ref{fig:tree} can be trained faster and with fewer costly global aggregations.

\subsection{Key Innovations in Fog Learning}
\label{ssec:innovation}
The key innovations of fog learning are as follows:
\begin{itemize}
\item Establishing multi-stage hierarchical machine learning through space.

\item Migrating from star to distributed learning topologies via collaboration/cooperation among D2D-enabled devices.

\item Employing agile network-aware management of heterogeneous nodes and channels.

\item Distributing task processing based on multi-objective network optimization of latency, cost, and privacy metrics.

\item Parallel successive learning for rapid refinement of locally trained models.
\end{itemize} 
 
\section{Open Research Directions}
\noindent In the following, we outline several directions of future research for fog learning:
 
\subsubsection{Optimizing horizontal/vertical communications}\label{sub1-1} 
Performing aggregations via D2D communications may be more resource-efficient, but can also incur more delay compared with the case of vertical aggregations. This delay is a function of data rates among the devices, channel qualities, rounds of D2D communication required, and other factors. 
Given the benefits of D2D communications discussed in Sec.~\ref{sec:adv}, quantifying the trade-offs and deciding which clusters of devices are suitable to perform the D2D communications deserves further investigation. Also, the potential for model inversion attacks at different network layers caused by horizontal parameter sharing needs to be considered, through effective countermeasures such as functional encryption~\cite{hardy2017private}.

\subsubsection{Multi-layer control and resource allocation}
Fog learning calls for a series of studies on designing control algorithms for orchestrating the nodes at different layers of the network. Along this direction, straggler mitigation in a multi-layer structure must be considered, along with asynchronous management of model aggregations. Additionally, efficient resource allocation along the cloud-to-things continuum must be considered, including congestion-aware distributed flow (load) balancing designs for distributed ML task handling. This may include a dynamic main server selection scheme based on network path resource availability.

\subsubsection{Error propagation analysis}
Due to communication imperfections and time-varying network topologies, horizontal parameter aggregations of devices in clusters may be noisy versions of the true aggregated values. Such noise will then be propagated and potentially amplified in transmission to upper layers. Modeling these errors, their propagation, and their cumulative effect on training convergence speed and accuracy is an interesting future direction.

\subsubsection{Intelligent cluster sampling}
To reduce power consumption and network traffic, the main server in Fig.~\ref{fig:tree} can perform cluster sampling, in which only the end devices from certain clusters engage in model training in each round. This requires considering end devices' data qualities and the characteristics of nodes in different network layers. Also, if nodes in the upper layers have mobile capabilities, this motivates network reconfiguration between global aggregations. For instance, instead of deploying a dedicated set of UAVs for data collection from each cluster of devices, a limited set of UAVs can be utilized, and the optimal trajectory can be obtained to enable the desired cluster sampling.

\subsubsection{Block-based learning}
\label{sub1-2}
The devices located in different layers of the network can form different \textit{learning blocks} (see Fig.~\ref{fig:tree}) to further decrease the network traffic and the required number of global aggregations.
In each block, the head (top-most) node(s) have a certain frequency of vertical communication. In-between vertical updates, they can conduct multiple rounds of in-block learning local updates. Studying the trade-offs between the aggregation frequencies of different learning blocks, the computational capabilities of the nodes inside the blocks, model accuracy, and training convergence speed is an open direction.

\subsubsection{Modeling of heterogeneous fog networks}
\label{sub1-3}
A comprehensive model of the interplay between fog network parameters (e.g., trust levels between users, D2D channel qualities, vertical communication quality variations, heterogeneous data quality, and heterogeneous compute capabilities) can lead to further optimization of fog learning.
Codifying each of these parameters and designing corresponding collaborative/cooperative learning schemes is an open direction.

\subsubsection{Smart data sharing}
End users can offload different parts of their datasets to different peers. In acting as helper nodes, devices with higher compute powers can send out requests for specific samples in a dataset that they lack 
to maximize the resulting data processing benefit. A similar procedure can be carried out using active inter-layer data caching. This will increase the quality of devices' datasets and improve the resulting global models.

\subsubsection{Incentivizing end users}
Proper incentive mechanisms are needed to persuade devices to participate in collaborative/cooperative model training. The incentives should consider the parameters of the local datasets (e.g., data quality) and the device's network-related parameters (e.g., speed of data offloading and computational capabilities). 
 
\subsubsection{Personalized model training} Training a single global model for an application can lead to poor performance at individual devices in scenarios of extreme data heterogeneity among geographically-distributed nodes. To address this, personalized model training can be investigated for fog learning through frameworks such as multi-task learning~\cite{8865093}.

\subsubsection{Dynamic networks and mobility models}
\label{sub1-4}
D2D data offloading and parameter sharing is only practical when mobile devices are within a certain vicinity. Accurate mobility models of devices could reveal pertinent information regarding the anticipated duration/frequency of contact, the data distributions of the contacting devices, and so forth. This information could be used to develop mobility-aware collaborative model training.

\subsubsection{Integration with wireless technologies}
Massive MIMO and reconfigurable intelligent surfaces are two radio technologies that will be major drivers of 5G-and-beyond wireless~\cite{TobeAdded9}. These physical/link-layer technologies can be developed jointly with fog learning to conduct model training over large numbers of users with high data rates and low latency.

\subsubsection{Deep reinforcement learning (DRL) for/via fog learning}
DRL  is a useful ML technique when perfect knowledge about the learning environment is not attainable \cite{luong2019applications}. This method has the potential to address design problems for wireless learning such as device beamforming, power control, interference management, coordination, and transmission scheduling, all of which can be adapted at different network layers. Decentralized training of DRL in turn requires message passing among the devices, which can be enabled at scale through fog learning via device collaboration, synchronization, and orchestration at different layers.

\balance

\section{Conclusion}
\noindent We introduced \textit{fog learning}, a new paradigm for distributing ML model training through large-scale networks of heterogeneous devices. We demonstrated that fog learning is inherently a multi-layer collaborative/cooperative hierarchical learning framework that can significantly reduce network resource costs and model training times through local model aggregations at different network layers. We introduced the hybrid property of fog learning, which combines horizontal D2D communications between nodes with vertical communications up the hierarchy. Further, we discussed the distributed topology and multi-objective optimization nature of fog learning that make it network-aware. Finally, we identified several open research directions in this emerging area.

\bibliographystyle{IEEEtran}
\bibliography{MagRefs}

\begin{IEEEbiographynophoto}{Seyyedali Hosseinalipour (M'20)} received B.S. degree from Amirkabir University of Technology in 2015 and Ph.D. degree from NC State University in 2020, both in electrical engineering. He received ECE doctoral scholar of the year award at NC State. He is currently a postdoctoral researcher at Purdue University. His research interests mainly include analysis of modern wireless networks and communication systems.
\end{IEEEbiographynophoto}
\begin{IEEEbiographynophoto}{Christopher G. Brinton (SM'20)}
is an Assistant Professor of ECE at Purdue University. His research interest is at the intersection of network optimization and data science. Since joining Purdue in 2019, he has won a seed for success award and an outstanding faculty mentoring award. He received his Masters and PhD in Electrical Engineering from Princeton University in 2013 and 2016, respectively, where he won the Bede Liu Best Dissertation Award. He is a co-founder of Zoomi Inc., and a co-author of the book \textit{The Power of Networks: 6 Principles that Connect our Lives}.
\end{IEEEbiographynophoto}
\vspace{-0.15in}
\begin{IEEEbiographynophoto}{Vaneet Aggarwal (SM'15)} received the B.Tech. degree in 2005 from the Indian Institute of Technology Kanpur, and the M.A. and Ph.D. degrees in 2007 and 2010, respectively from Princeton University, all in Electrical Engineering. He is currently an Associate Professor at Purdue University. 	He received Princeton University's Porter Ogden Jacobus Honorific Fellowship in 2009, the 2017 IEEE Jack Neubauer Memorial Award, and the 2018 Infocom Workshop Best-Paper Award. His current research interests are in communications and networking, cloud computing, and machine learning.
\end{IEEEbiographynophoto}
\vspace{-0.15in}
\begin{IEEEbiographynophoto}{Huaiyu Dai (F’17)}
received the B.E. and M.S. degrees in electrical engineering from Tsinghua
University, Beijing, China, in 1996 and 1998, respectively, and the Ph.D. degree in electrical
engineering from Princeton University, Princeton, NJ in 2002. He is currently a Professor of Electrical
and Computer Engineering with NC State University, Raleigh, holding the title of University Faculty
Scholar. His research interests are in the general areas of communication systems and networks,
advanced signal processing for digital communications, communication theory, and information
theory.
\end{IEEEbiographynophoto}
\vspace{-0.15in}
\begin{IEEEbiographynophoto}{Mung Chiang (F'12)} is the John A. Edwardson Dean of the College of Engineering at Purdue University. He received his B.S. (Honors), M.S. and Ph.D. from Stanford University in 1999, 2000, and 2003 respectively. Prior to coming to Purdue, he was the Arthur LeGrand Doty Professor of Electrical Engineering at Princeton University. His research on networking received the 2013 Alan T. Waterman Award, the highest honor to US young scientists and engineers. His textbook \textit{Networked Life} and online course reached 250,000 students since 2012, and the popular science book \textit{The Power of Networks} was published in 2016. He founded the Princeton EDGE Lab in 2009, which bridges the theory-practice gap in edge networking research by spanning from proofs to prototypes. He co-founded startups in mobile, IoT and big data areas, and co-founded the global nonprofit Open Fog Consortium.
\end{IEEEbiographynophoto}
\end{document}